\documentclass[12pt,letterpaper,oneside]{article}
\usepackage[numbers]{natbib}
\usepackage{graphicx}
\usepackage[applemac]{inputenc}
\usepackage{fancyhdr}
\usepackage{courier}
\usepackage{float}
\usepackage{geometry}
\usepackage{epsfig}
\usepackage{color}
\usepackage{makeidx}
\usepackage{eurosym}
\usepackage{rotating}
\usepackage[stable]{footmisc}
\usepackage{multicol}
\usepackage{epsfig}
\usepackage{caption}

\def\mbh{$M_{\rm BH}$\/}

\def\lledd{$L/L_{\rm Edd}$}

\def\rfe{$R_{\rm FeII}$}

\def\feiiq{{\rm Fe}{\sc ii}$\lambda$4570\/}
\def\msol{M$_\odot$\/}

\def\rg{$r_{\rm g}$\/}
\def\ltsima{$\; \buildrel < \over \sim \;$}
\def\ltsim{\lower.5ex\hbox{\ltsima}}  
\def\gtsima{$\; \buildrel > \over \sim \;$}

\def\gtsim{\lower.5ex\hbox{\gtsima}}

\def\lya{{ Ly}$\alpha$}
\def\civ{{\sc{Civ}}$\lambda$1549\/}

\def\cm3{cm$^{-3}$\/}
\def\hb{{\sc{H}}$\beta$\/}

\def\mgii{{Mg\sc{ii}}$\lambda$2800\/}

\def\ciii{{\sc{Ciii]}}$\lambda$1909\/}
\def\oiiiopt{{\sc{[Oiii]}}\-$\lambda\lambda$\-4959,\-5007\/}
\def\o4363{{\sc{[Oiii]}}$\lambda$4363\/}

\def\siiii{Si{\sc iii]}$\lambda$1892\/}
\def\aliii{Al{\sc iii}$\lambda$1860\/}

\def\feiiopt{{Fe \sc{ii}}$_{\rm opt}$\/}
\def\feii{{Fe\sc{ii}}\/}

\def\feiii{{Fe\sc{iii}}\/}
\def\fe{{\sc{Fe}}\/}
\def\gs{{$\Gamma_{\mathrm{soft}}$\/}}

\def\fe76087{{\sc [Fe vii]}$\lambda$6087\/}

\def\kms{km~s$^{-1}$}

\def\ergss{ergs s$^{-1}$\/}

\def\apj{ApJ}
\def\apjl{ApJL}
\def\apjs{ApJS}
\def\pasj{PASJ}
\def\aj{AJ}
\def\apss{ApSpSci}
\def\araa{AnnRevAAp}
\def\mnras{MNRAS}\def\pasp{PASP}
\def\aap{AAp}
\def\nat{Nat}

\def\siiv{Si{\sc iv}$\lambda$1397\/}

\def\om{$\Omega_\mathrm{M}$}
\def\ol{$\Omega_{\Lambda}$}

\usepackage{titlesec}
\usepackage{tocloft}
\title{\bf\sc Low- and high-$z$ highly accreting quasars\\ 
in the 4D Eigenvector 1 context\footnote{Submitted to the {\em Astronomical Review}.}}
\author{Paola Marziani\footnote{INAF, Osservatorio Astronomico di Padova, Italia.}, Jack W. Sulentic\footnote{Instituto de  Astrof\'{\i}sica  de Andaluc\'{\i}a (CSIC), Granada, Spain.}$^{~\parallel}$,  C. Alenka Negrete\footnote{INAOE, Puebla, M\'exico.}, \\    Deborah Dultzin\footnote{Instituto de  Astronom\'{\i}a, UNAM, M\'exico.}, Mauro D' Onofrio\footnote{Dipartimento di Fisica e Astronomia ``G. Galilei,'' Università di Padova, Italia.}, Ascensi\'on del Olmo$^{\ddag}$,\\ and Mary Loli Mart\'{\i}nez-Aldama$^\P$\\}
\date{}
\begin{document}
\maketitle

\begin{abstract}
Highly accreting quasars are characterized by distinguishing properties in the 4D eigenvector 1 parameter space  that make them easily recognizable over a broad range range of redshift and luminosity.  The 4D eigenvector 1 approach allows us to define   selection criteria that go beyond the restriction to Narrow Line Seyfert 1s identified at low redshift. These criteria are probably able to isolate sources  with a defined physical structure i.e.,  a geometrically thick, optically thick advection-dominated accretion disk  (a ``slim'' disk).  We stress that the importance of highly accreting quasars  goes beyond the understanding of the details of their physics: their Eddington ratio is expected to saturate toward values of order unity, making them possible cosmological probes.
\end{abstract}

\section{Introduction}


It is now widely accepted that quasars are a manifestation of accretion over a massive compact object, most likely a black hole (see \citep{donofrioetal12,netzer13} for reviews and for an excellent systematic presentation). The gravitational force exerted by a black hole on the surrounding gas is of course proportional to its mass  \mbh.  Accretion phenomena  are however rather mass invariant, so that no great observational diversity associated with black hole mass is expected: accretion luminosity can be written as $L_\mathrm{acc} \propto M_\mathrm{BH} \dot{M} / r$,  where $r$\ indicates a distance from the  black hole customarily identified  with the radius of the innermost stable close orbit (ISCO). The $r_\mathrm{ISCO}$\ is a decreasing function $f(a)$ of the black hole specific angular momentum per unit mass $a$,  but in any case  $\sim$ \rg. It follows that  $L_\mathrm{acc} \propto \dot{M} f(a)^{-1}$:  the energy output is independent of the black hole mass \mbh\ and explicitly dependent on $a$ \citep{pagethorne74,wangetal14}.  Not surprisingly, similarities have been  found between stellar mass  and supermassive black holes that include superluminal motion \citep{mirabelrodriguez94},   accretion disk structure (the standard $\alpha$- accretion disk model applied to quasar continuum  was originally  developed for black holes in binary stellar systems, \cite{shakurasunyaev73}),   presence of a hard- (low accretion) and soft- (high accretion) state \cite[e.g.,][]{poundsetal95,janiukczerny07} affecting the shape of the soft X-ray continuum \citep[e.g.,][]{wangetal96,grupe04}, possibly through a jet-accretion disk symbiosis \citep{falckebiermann99,kordingetal06}.\footnote{Powerful radio emission is unsteady, or even impossible  at high \lledd \citep{fenderbelloni04,pontietal12,neilsenlee09} for Galactic black hole candidates. It is still debated whether a close analogy can be made with quasars. }  An interesting analogy involves even the Narrow Line Seyfert 1 (NLSy1) prototype I Zw 1 and  two accreting white dwarfs  with jets \cite{zamanovmarziani02}, where the optical \feii\ emission appears to be strikingly similar. 


Type-1 Active Galactic Nuclei (AGN) and quasars   do however show a great diversity in observational properties \citep{sulenticetal00a}.  Physical parameters others than \mbh\ alone should be sought to explain their diversity.  Since quasars radiate up to $10^{48}$ \ergss, it is also reasonable to assume that a relevant force acting on the photoionized, line emitting gas at 100 -- 1000 \rg\ the black hole is due to radiation. Internal broad line shifts in the spectra of quasars  have turned out to be powerful diagnostics of radiation forces, and their exploitation has gone on a par with the availability of multi-frequency observations since the early 1980s \cite{gaskell82,brothertonetal96,marzianietal96}. High ionization lines (HILs, most notably \civ) have been found to show systematic blueshifts that can reach several thousands \kms\ with respect to the low-ionization lines (LILs) and to the best estimates of the rest frame of the host galaxy. Evidence of outflows is now considered overwhelming on small (1pc) and large (kpc) spatial scales \cite{everett07,marzianisulentic12a,crenshawetal10}. On small scales   compelling evidence is provided by   blueshifted emission lines \citep{gaskell82,marzianietal96}, blueshifted broad absorptions lines (BAL) observed in $\approx$ 10 \%\ of QSOs \citep[see][for a fully developed line of reasoning]{peterson97}, as well as by blueshifted  features belonging to the soft X warm absorber or to ``ultra-fast outflows'' \citep[e.g.,][]{blustinetal05,tombesietal13}.  These finding indicate that  radiation forces are   likely driving an high-ionization outflow from the accretion disk surrounding the central black hole \cite{netzer13}.  Several observational properties seem to be indeed governed by the forces due to gravity and radiation, and by their relative balance \citep[][]{marconietal09,ferlandetal09,netzermarziani10,marzianietal10}. 

The hope of organising quasar  observational diversity on a physical basis rests therefore on defining a parameter that is $\propto \dot{M}$\ normalized by \mbh.  The key to recognise highly accreting quasars is  to find  efficient observational markers that unambiguously identify them by exploiting effects of accretion and radiation forces.  Theoretically, there is a limit of $\approx$ a few times the Eddington luminosity \cite{mineshigeetal00} for sources accreting at an arbitrarily large dimensionless accretion rate $\dot{m} \gg 1$. Observationally, conventional Eddington ratio estimates  (\S \ref{mbh}) indicate an lower limit at \lledd $\sim$ 0.01\ and an upper limit at $\approx$ 1. Therefore the term highly accreting (xA) quasars  will be used as synonym for quasars accreting  close to the Eddington limit.\footnote{\lledd\ is a parameter that can be fully determined from observations: \lledd $\propto  L/$\mbh, the ratio of accretion (bolometric) luminosity to black hole mass. On the converse the dimensionless accretion rate is $\dot{m} = \dot{M}/\dot{M}_\mathrm{Edd}$, with $L = \eta \dot{M} c^2$. The efficiency $\eta$ can be factored into two terms: $\eta = \eta_1(\dot{M}) \eta_2(a)$. Since $\eta_{2}(a)$\ of an individual quasar is presumably the same for all accretion rates,  $ \dot{m} = L/L_\mathrm{Edd} \cdot \eta_1(\dot{M})/\eta_1(\dot{M}_\mathrm{Edd})$, which may imply $\dot{m} \gg L/L_\mathrm{Edd}$\   if $\eta_{1}$ decreases with increasing $\dot{M}$.}  

It is possible to empirically organise quasars following the so-called eigenvector 1 (E1) scheme  \citep{borosongreen92}, especially in its 4D formulation (4DE1: \cite{sulenticetal00a,sulenticetal00b}, \S \ref{4de1}).  The empirical organisation defines a quasar main sequence that is primarily a sequence of Eddington ratio \citep{marzianietal01,aietal10,marzianietal13}, as expected from the elementary considerations made above.  After reviewing the systematisation of quasar diversity,   the physical interpretation of eigenvector 1 related correlations (\S \ref{phys}) and the rationale for two quasar populations (\S \ref{popab}), we will turn to  the identification of extreme accretors  at low- and high-$z$\ .   Even if  quasar sample sizes  increased tremendously in the last decades, there are still major observational limitations related to an unavoidable Eddington ratio bias (\S \ref{ebias}). Overcoming this bias  requires  criteria that are as much as possible luminosity-independent  (\S \ref{xA}).  Last we  
discuss the possible  exploitation  of highly accreting sources as distance indicators (\S \ref{cosmo}).

\begin{figure}[htp!]
\begin{minipage}[t]{0.475\linewidth}
\centering
\includegraphics[width=3.1in]{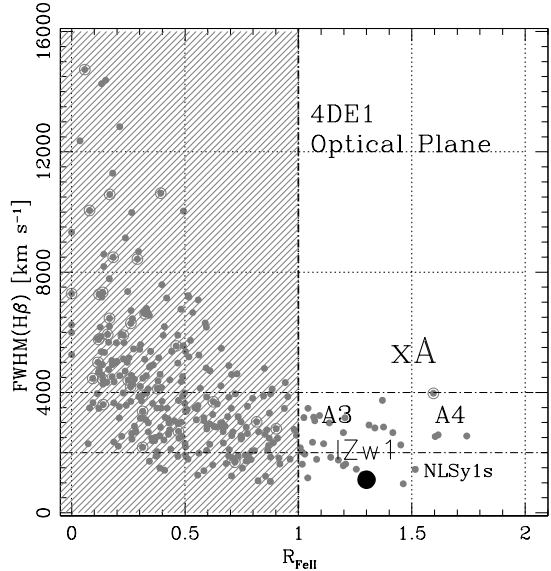}
\end{minipage}
\hspace{0.7cm}
\begin{minipage}[t]{0.475\linewidth}
\vspace{-7.cm}
\centering
\caption{The optical plane of the 4DE1 parameter space, FWHM (\hb) (broad component) vs. \rfe. Grey dots represent the data points of \cite{zamfiretal10} for $z \le 0.7$, with a high \rfe\ tail extending above \rfe=1, considered as a defining criterion for identifying higly accreting (xA) quasars. Note that part of Pop. A (actually, about 80\%\ of Pop. A quasars) are excluded (shaded area). The limit for NLSy1 and for Pop. A at low-$L$ ($\log L \ltsim 46$\ [\ergss]) is shown along with the position of the prototype source I Zw 1. }\label{fig:e1f}
 \end{minipage}
\end{figure}

\section{Organizing quasar diversity: the 4DE1 approach}
\label{4de1}

Over the past 15 years our group developed an empirical context to interpret the spectroscopic diversity of quasars \citep{marzianietal96,sulenticetal00a}. The work began at low redshift ($z < 0.8$) where a four-dimensional parameter space has been exploited using optical, UV and X-ray measures \citep{sulenticetal00b} to define a 1 dimensional sequence (the so-called Eigenvector 1 of quasars; \cite{marzianietal01}). The four dimensions of the parameter space are defined as follows:

\begin{itemize}
\item Full width at half maximum of broad \hb\ (FWHM \hb) which is our diagnostic low ionization line (LIL) accessible out to $z = 0.7$ with optical (and intermittently to $ z=3.7$\ with infrared) spectroscopy \citep{marzianietal09,zamfiretal10,popovickovacevic11,steinhardtsilverman13}. FWHM \hb,  FWHM \mgii, \cite{marzianietal13a,marzianietal13}, or perhaps even FWHM Pa$\alpha$\ \cite{lafrancaetal14} are thought to be a measure of   virialized motions in the line emitting regions and are crucial for estimating \mbh.
\item Intensity	ratio of  the 4570 \AA\ optical \feii\ blend and broad \hb\ (\rfe  =   \feiiq/  \hb ). \rfe\ is sensitive to the ionization conditions and column density of the BLR gas \citep{marzianietal01,ferlandetal09,marzianietal10} arising, at least to a first approximation, in some of the gas that produces \hb.
\item Centroid shift at half maximum of the HIL \civ\ ($c(\frac{1}{2})$). It is a diagnostic of winds/outflows \cite{marzianietal96,richardsetal02}.
\item 	Soft X-ray photon index (\gs). \gs\ is diagnostic of the thermal emission from the accretion disk, sensitive to the accretion state \citep{wangetal96,grupe04},   
\end{itemize}

It is important to stress that the Eigenvector 1 correlated parameters  are many more than the ones involved in the 4DE1 space: among them,   the spectral energy distribution \citep{kuraszkiewiczetal09,tangetal12}, the relation between 
  X-ray continuum and \civ\ properties \citep{kruzceketal11}, the ratio between optical and UV \feii\ emission \citep{sameshimaetal11}, and most notably,  the ratios \aliii / \siiii, and \siiii/ \ciii\ \citep{baldwinetal96,willsetal99,bachevetal04,marzianisulentic14} discussed in \S \ref{xA}. The E1 space extends also to the optical variability of quasars \citep{maoetal09}, the far-IR luminosity \cite{wangetal06}. Even the Baldwin Effect -- hailed as the most important luminosity effect -- is at least in part due to the HIL  equivalent width correlations associated with E1  \citep{bachevetal04,baskinlaor05b,marzianietal06,marzianietal08}. 

Fig. \ref{fig:e1f} shows the optical plane of the 4DE1 parameter space, FWHM (\hb) (broad component) vs. \rfe, with lines separating Pop. A and B and NLSy1s sources. The dotted lines identify spectral bins in the plane separated by $\Delta$\rfe =0.5 and $\Delta$ FWHM\hb = 4000 \kms\ (\S \ref{popab}). The eigenvector 1 results make it possible, once an empirical sequence is established, to consider two approaches in data analysis: (1) the use of a limited number of sources (few, or even 1 at worst, \cite{marzianietal10}) that belong to the same spectral bin in the sequence and that therefore show consistent optical and UV properties; (2) the use of larger samples, especially if the individual data are of low S/N. In this case, median composites can be meaningfully built for individual sources belonging to the same bin along the sequence, if and only if individual data S/N is high enough to allow for a bin assignment.  This minimum S/N is a requirement to avoid systematic effects on FWHM and EW of lines (see, for example, Fig. 5 of \cite{shenetal11}). UV and optical composite spectra along the E1 sequence have been shown by \citep{bachevetal04,marzianietal09}.

\section{Physical  interpretation of E1}
\label{phys}

\subsection{Black hole mass and Eddington ratio computations}
\label{mbh}


The virial black hole mass can be written as $M_\mathrm{BH} = f_\mathrm{S} r_\mathrm{em} \delta v^2 / G$, where $f_\mathrm{S}$\ is the emitting region structure factor,    $\delta v$\ is the virial broadening estimator, and $r_\mathrm{em}$\ is a characteristic distance from the black hole of the line emitting gas (i.e., in practice the distance derived from reverberation mapping, \cite{shen13,marzianisulentic12a} for reviews). Evidence in support of virial motion for the line emitting gas comes from three lines of investigation: (1) velocity resolved reverberation mapping studies \citep{gaskell88,koratkargaskell91,grieretal13}  that exclude  outflow as the broadening source in at least LILs; (2) the anticorrelation  between line broadening and the time lag of different lines in response to continuum variation \citep{petersonwandel99,petersonetal04,krolik01}. This   relation has been found for a few nearby objects extensively monitored (none of them an extreme radiator) with a slope close to the one expected from the virial relation.   (3) More circumstantial evidence for virial motion is provided by symmetric and unshifted\footnote{Within 100-200 \kms\ from rest frame.}  \hb\ line profiles of NLSy1-like sources  \citep{veroncettyetal01,marzianietal03a}.  There is a growing consensus that the line width of LILs (\hb\ and \mgii) can be considered a reliable virial broadening estimator \cite{marzianietal13a,trakhtenbrotnetzer12}.  UV intermediate emission lines -- but not an HIL like \civ! \citep{sulenticetal06,netzeretal07} -- have been found suitable as well \citep{negreteetal13}, at least for low-$z$ quasars.  

A flattened geometry of the line emitting region \cite{willsbrowne86,mclurejarvis02,collinetal06,kollatschnyzetzl11} is suggested by the dependence of line width on the viewing angle. There are  few fortunate cases when the viewing angle can be estimated (for example, if superluminal motion is detected: \citep{sulenticetal03}); otherwise, orientation effects emerge  on a statistical basis by comparing the average line widths of LD and CD RL sources \citep{marzianietal03b,zamfiretal08}. Accurate and precise measures of the quasar  luminosity and the connection between radial velocity and deprojected velocity are orientation dependent \citep{runnoeetal13}.  However, the viewing angle is a parameter that is in general unknown for all radio-quiet sources, both of Pop. A and B. 

Computation of Eddington ratio requires an estimate of a bolometric correction that is most likely orientation- and possibly luminosity-dependent. Assuming a constant bolometric correction from the optical and UV does not seem   appropriate even if the scatter around an average may not be very large \cite{richardsetal06}. The bolometric correction is certainly dependent upon E1, although a systematic estimate of the correction as a function of spectral type for optical and UV observations is not available as yet. 




\subsection{The 4DE1 main sequence }

The 4DE1 diagram  can be seen as a surrogate H-R diagram for quasars, and is now understood to be mainly driven by Eddington ratio \citep{sulenticetal00a,marzianietal01,marzianietal10,boroson03,ferlandetal09,dongetal09}. There is a systematic Eddington ratio trend along the quasar main sequence  in Fig. \ref{fig:e1f},  with Pop. A sources being  higher radiators.  The trend is well illustrated in a study where  spectral types are defined over a  sample of 680  Sloan Digital Sky Survey (SDSS) sources, and \lledd\ is computed for each spectral type \citep{marzianietal13a}. $\log$ \lledd\ increases monotonically  along the 4DE1 sequence  from --1.62 (spectral type B1$^{++}$)  to --0.14 (spectral type A4 representative of xA sources). 

The blue ward asymmetry observed in the \civ\ profile indicates that at least a part of the HILs is emitted in outflowing gas where the receding part of the flow is obscured \citep{gaskell82,marzianietal96,richardsetal06}.  The prominence of the outflow is maximised at the high Eddington ratio end of the 4DE1 sequence   \citep{marzianietal96,marzianietal10}.  

Orientation  is acting as a source of scatter smearing the main sequence in the 4DE1 optical plane \citep{marzianietal01,sulenticetal03}.    Black hole mass  and, to a lower extent, metallicity are also found to have a broadening effect on the 4DE1 sequence \citep{sulenticetal01,zamanovmarziani02}. We have also to consider that, even if 4DE1 eases the systematization of quasar spectral properties, a full understanding of some  parameters used in the formulation of the 4DE1 itself is still missing. For example, the  emission mechanism of \feii\ is still unclear, especially for strong \feii\ emitters as found in A3 and A4 spectral types   (\citep{collinjoly00,jolyetal08,marzianietal13b} review of the ``\feii\ problem").

\section{Two AGN populations: A and B, and the standard model}
\label{popab}

The first suggestion  of two  populations of unobscured type 1 quasars and luminous Seyfert 1 nuclei was made in 2000 \citep{sulenticetal00a,sulenticetal00b}, on the basis of the source distribution in    4DE1. The two populations (A and B(roader)) were separated at the empirical limit FWHM \hb $= 4000$ \kms\ that can be deduced from Fig. \ref{fig:e1f}.  Population A and B are interpreted as populations of sources accreting at different rates \citep{marzianietal03b}, with higher rate  for Pop. A (that includes classical NLSy1s) and lower for Pop. B (that includes the most evolved FRII radio sources; see reviews on the topic by \citep{sulenticetal08,sulenticetal11}; \citet{sulenticetal07} provide a table listing major differences between the two Populations).  Here we remind that Pop. A sources show large \gs,   a high frequency of large \civ\ blueshifts \citep{sulenticetal07}, as well as low  \civ\ EW (typically $\approx$50 \AA\  \citep{sulenticetal07}). The A/B separation has been recognised by several workers, who discuss   \hb\ profile shape differences for narrower and broader sources \citep{collinetal06,marzianietal03b,sulenticetal02,zamfiretal10}.  Pop. A sources with   blue shifts have been called  wind dominated sources, and Pop. B   disk dominated  \citep{richardsetal11}. A soft state (i.e., large \gs) with  strong winds has been identified in quasars and stellar mass black holes as well \citep{pontietal12}.

\begin{figure}[ht]
\begin{minipage}[t]{0.65\linewidth}
\centering
\includegraphics[width=4.5in]{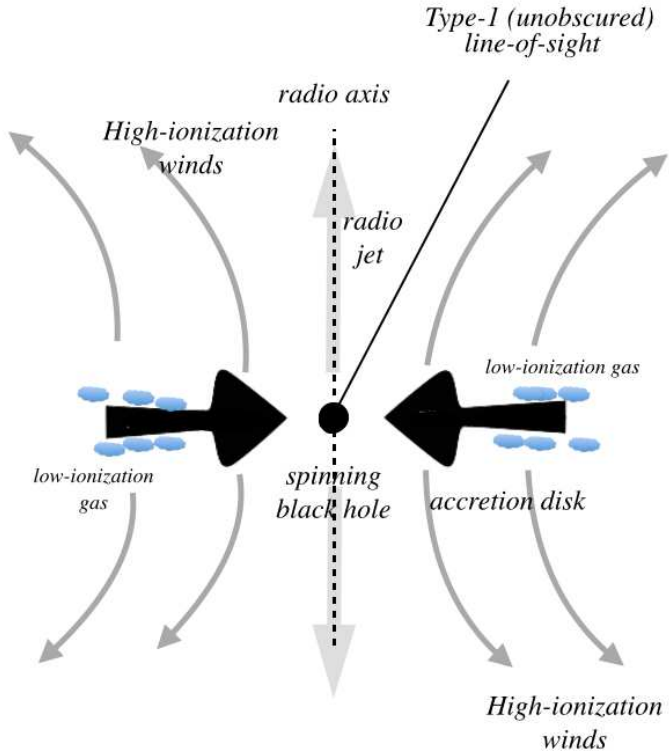}
\end{minipage}
\hspace{0.7cm}
\begin{minipage}[t]{0.3\linewidth}
\vspace{-9cm}
\centering
\caption{A scheme illustrating the main physical constituent of Pop. A and xA nuclei, with a geometrically thick advection flow around a central spinning black hole that defines a preferential symmetry plane. The line of sight is shown  at the most likely inclination with respect to the disk axis ($\approx$ 30$^\circ$). Inclinations larger than 45$^\circ$ -- 60$^\circ$ are expected to intercept an obscuring torus (not shown) and suppress a direct view of the broad emission lines.   }\label{fig:stru}
 \end{minipage}
\end{figure}

Conservation of angular momentum leads to the formation of a Keplerian (or quasi-Keplerian) accretion disk around the center of gravity provided by the black hole. Over the years, optical and UV continuum fits have revealed the stretched black body   spectral energy distribution (SED) that supports  the accretion disk idea \cite{shields78,malkansargent82}.\footnote{Perhaps one of the most significant  paradigm shift in AGN research concerned  the interpretation of the optical/UV continuum of luminous quasars: the ``featureless continuum"  is now ascribed to the low-energy tail of thermal emission from an accretion disk, not anymore to non-thermal emission.}  Theoretical models indicate  that the disk is optically thick within the range of accretion rates of type-1 quasars ($\dot{m} \gtsim$ 0.01, \citep{marzianietal03b,narayanyi94}). Geometrical thickness is  affected by accretion rate: the accretion disk is expected to form a ``slim'' structure if $\dot{m}\gtsim$ 0.1-0.3 \cite[e.g.,][]{abramowiczetal88}. This condition is consistent with an \lledd\ limit that was derived from observations $\approx 0.15$\ \citep{marzianietal03b}. 
Fig. \ref{fig:stru} shows a scheme illustrating a possible structure for   quasars of Pop. A. The main physical constituent  are a geometrically thick advection flow around a central spinning black hole that defines a preferential symmetry plane. LIL emitting gas may be in a flattened cloud configuration  associated with the disk at the base of high-ionization winds.  At low accretion rates, the flow may remain geometrically thin down to $r_\mathrm{ISCO}$.  

A recent, extensive   review on accretion disks is provided by \citet{abramowiczstraub14}. Here we point that that the rapid radial drift due to the efficient loss of angular momentum  in the advection dominated accretion flow makes it difficult to predict structure and observational properties.  There is still no clear, complete  model connecting the slim disk structure to the observed emission line profiles for LILs, and to the strong high ionization wind present in sources where a slim disk is expected. Only a  very recent   work attempts to explain the observed  Lorentzian profiles in Pop. A as  due to   two components, one exposed to the hottest part of the slim disk within a funnel, and and the other  affected by self-shadowing \citep{wangetal14b}.  Self-shadowing and anisotropy in continuum emission should be considered especially in light of the possible cosmological exploitation of extreme accretors   (\S \ref{cosmo}, \citep{marzianisulentic14,wangetal13,wangetal14b}).  However, the distinction between Pop. A and B is the first step to recognise highly accreting quasars as the Pop. A / B transition  be associated with a change in structure of the accretion flow. 

\section{A Universe of quasars}

There has been a tremendous growth in data availability for quasar that makes even more necessary a systematic organisation of their properties in physical terms. We have moved from the color-based selection of quasars of the 1980s that yielded samples of  $\sim$ 100 objects with ``good" spectra, to progressively larger samples like the one of the LBQS that, in the early 1990s, involved $\sim$ 10$^3$ sources. The early data release of the SDSS accounted for another order of magnitude leap in sample size, with $\sim 10^4$ sources.  The IIIrd SDSS generation has lead to several $10^5$\ confirmed quasar, with the SDSS  IV-BOSS (Baryon Oscillations Sky Survey)  yielding candidate sources now exceeding one million \citep{parisetal14,flesch13}. The optical surveys have been paired with matching X-ray surveys:  the XMM-Newton serendipitous survey \citep{watsonetal09} reaches  flux limits typically in the range $1 - 5  \times 10^{-15}$ \ergss\ cm$^{-2}$ nicely complement some existing SDSS data for  $i < 22$ type 1 quasars at $z = 2$. FIRST \cite{beckeretal95} and Herschel-Atlas \cite{ealesetal10} data also provide radio and sub-mm counterparts for optically selected AGN over a large fraction of the celestial sphere.  How will these new and less new data be organized? Will all sources be confused together with the naïve expectation that all quasars are the same object? This attitude – that was stressed   as the foundation of quasar studies by an influential speaker at a recent international meeting -- has proved to be confusing, suffocating, given the great diversity in quasar spectral properties. Current practice however does not go beyond a generic distinction between type 1 and 2, and between RL and RQ sources.  4DE1 helped recognise important trends at low-$z$. Its extension at higher $z$\ requires some care
with the consideration of  \mbh\ and luminosity effects that are weak but become significant if  \mbh\ and luminosity span several decades.

\begin{figure}[t]
\centering
\includegraphics[width=3.1in]{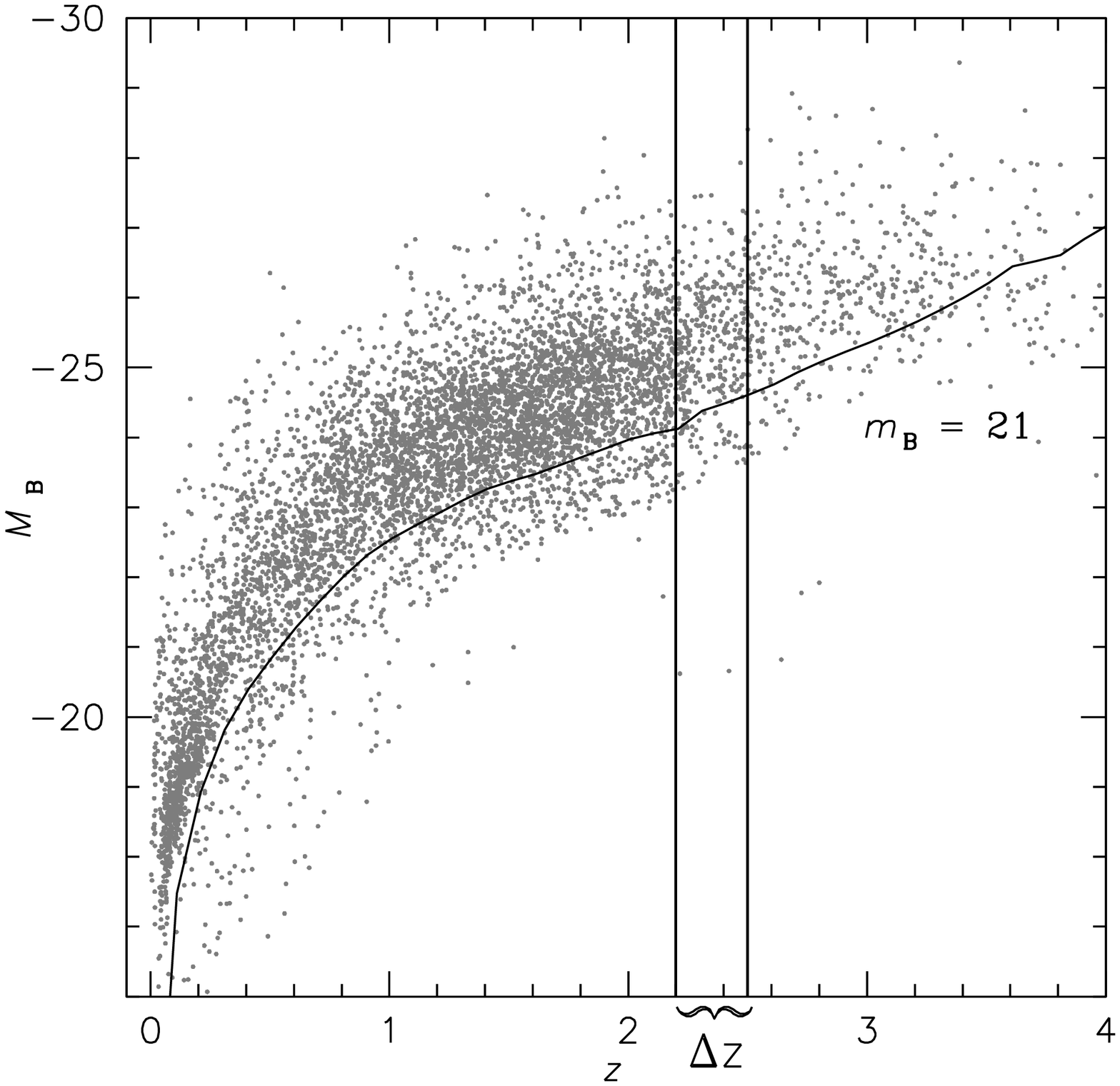}
\includegraphics[width=3.1in]{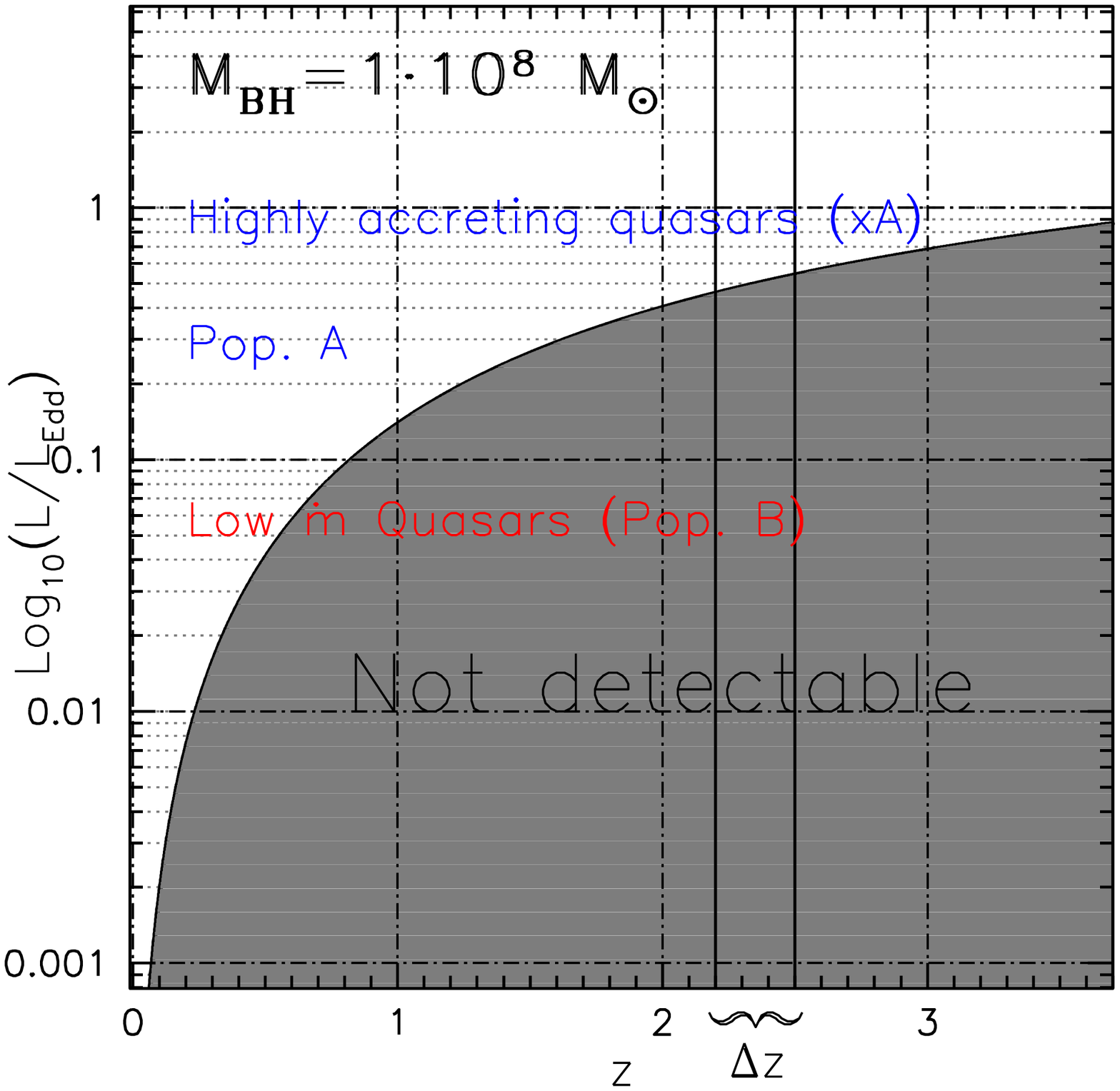}
\caption{Left panel: Effect of the introduction of a defined flux limit    at $m_\mathrm{B} = 21$   (curved black line) in the parameter plane absolute B magnitude $M_\mathrm{B}$ vs. redshift $z$. Data are derived from the catalog of \cite{schneideretal10} based on SDSS data release (DR) 7. The lines  at  $z \approx 2.2 $ and $z \approx 2.5$\ trace the redshift limits of a hypothetical pencil beam survey as the one carried out by Sulentic et al. \citep{sulenticetal14a}. Note that even with a relatively deep limiting magnitudes, all AGN of Seyfert-level luminosity would be lost. Right panel: ``Eddington ratio bias", shown as function of redshift for a fixed black hole mass ($10^8$\msol). The shaded area defines the \lledd\ domain where sources would not be detectable for the same limit at $m_\mathrm{B} = 21$\ as in the left panel.  Highly accreting quasars (\lledd $\approx$ 1, extreme Population A) will remain detectable up to very high $z$, while there will be a progressively increasing loss of Pop. A sources radiating at \lledd $\gtsim 0.2 $. All  quasar accreting at low or modest Eddington ratio (Pop. B),  $0.01 \ltsim $ \lledd $\ltsim 0.2$ are lost at $z \gtsim 1$, and therefore also in a pencil beam survey between  $z \approx 2.2 $ and $z \approx 2.5$. }
\label{fig:bias}
\end{figure}

\subsection{The Eddington ratio bias\footnote{The wording ``Eddington ratio bias" should not be confused with the  ``Eddington bias"  \cite{eddington13}, i.e., the bias introduced by random error in magnitude measurements on number counts of sources that leads to an overestimate of the number of faint sources. We intend here a \mbh\ and $z$- dependent loss of sources introduced by a flux limit.}}
\label{ebias}

We usually have no synoptic view of a broad wavelength range for individual quasars.  Accessible rest-frame ranges provided by survey spectra depend on $z$:
\hb:      $0 < z \ltsim 0.7$;        \lya\       $z \gtsim 2.0 $ ;    \civ:   $z \gtsim 1.4$; \siiv;    $ z \gtsim 1.7    $;         1900 blend (\aliii + \siiii + \ciii + \feiii):  $ z\gtsim1.1$. Since LILs are mainly in the optical and HILs in the UV,  our ability 
to perform the usually fruitful comparison between LIL and HIL profiles  is still limited to small samples of $\sim$ 100 sources.
                     
In addition, the flux limits associated with optical surveys cause what is literally an ``Eddington ratio bias." Sources with larger masses will be detected radiating over a broader range of Eddington ratios than sources with smaller masses. Actually, typical black hole masses of local NLSy1s would go undetected at $z  \approx 2$\ at the flux limits of the deepest currently available surveys (Fig. \ref{fig:bias} and Fig. 2  of \cite{sulenticetal14a}).  
We are forced to study faint quasars at low $z$, and luminous quasars at high $z$. If we define as intrinsically faint a quasar whose luminosity is below the knee of the quasar luminosity function \cite{boyleetal00,richardsetal06}, those quasars are in the high luminosity tail of the local quasar luminosity function. We do not observe active black holes of large mass ($\gtsim 10^{9}$ \msol) in the local Universe. On the other hand, even the deepest surveys do not sample -- or are at least biased against -- the small mass black holes that are observed in local quasars. Therefore, when we consider mass effects we have to compare sources that are at very different redshift. This makes very hard or even impossible to distinguish between true luminosity and evolutionary ($z$-dependent) effects.

To recognise highly accreting quasars at low- and high-$z$ we have   to define criteria that are not strongly luminosity dependent. If we resort to optical and UV rest frame spectra (still major identification tools, and going to stay so for a foreseeable time) then we have to find equivalent criteria for optical and UV data that are equivalent in finding highly accreting quasars. 

The Eddington ratio bias should not hamper our ability to find large \mbh\ quasars radiating at high \lledd\ even at the highest $z$.  Small \mbh\ quasars, even if radiating close to the Eddington limit, are however not detected at high redshift even from the deepest quasar surveys: they are simply not reported in present-day catalogues.  This is made explicit by the left panel of Fig.   \ref{fig:bias} illustrates the selection effects in the redshift absolute magnitude plane by the flux limit of the SDSS. The right panel of  Fig. \ref{fig:bias} shows the concept of Eddington ratio bias for a fixed black hole mass: there is $z$-dependent limit to sources with detectable \lledd.


\begin{figure}[ht]
\begin{minipage}[t]{0.475\linewidth}
\centering
\includegraphics[width=3.1in]{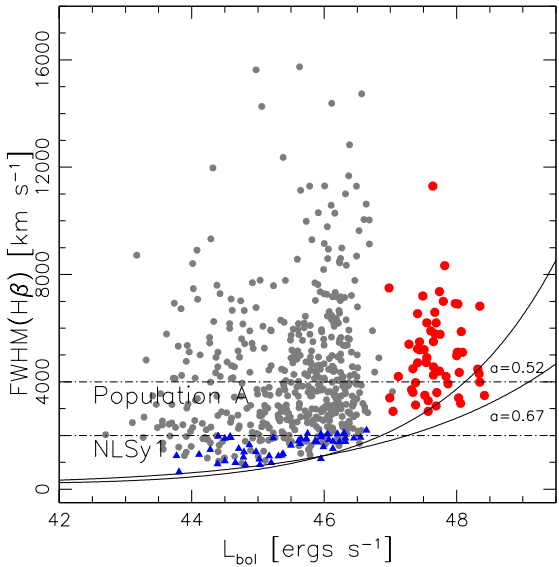}
\end{minipage}
\hspace{0.7cm}
\begin{minipage}[t]{0.475\linewidth}
\vspace{-7.cm}
\centering
\caption{Luminosity effects on FWHM(\hb), shown against $\log L$ for the samples of \cite{zamfiretal10} (grey dots), \cite{zhouetal06} (blue dots), and the high $L$\ sample of \cite{marzianietal09}  (red dots). The two lines show the minimum FWHM as a function of luminosity for sources radiating at Eddington limit, assuming the \mbh\ can be estimated from the virial relation, and that there is a Kaspi-type power-law scaling relation for the emitting region size with luminosity  ($r \propto L^a$). Two cases, for $a \approx 0.52$\ and $a \approx 0.67$, are shown. \label{fig:lum}}
\end{minipage}
\end{figure}

\section{Highly  accreting Pop. A sources}
\label{xA}

\begin{figure}[t]
\centering
\includegraphics[width=3.1in]{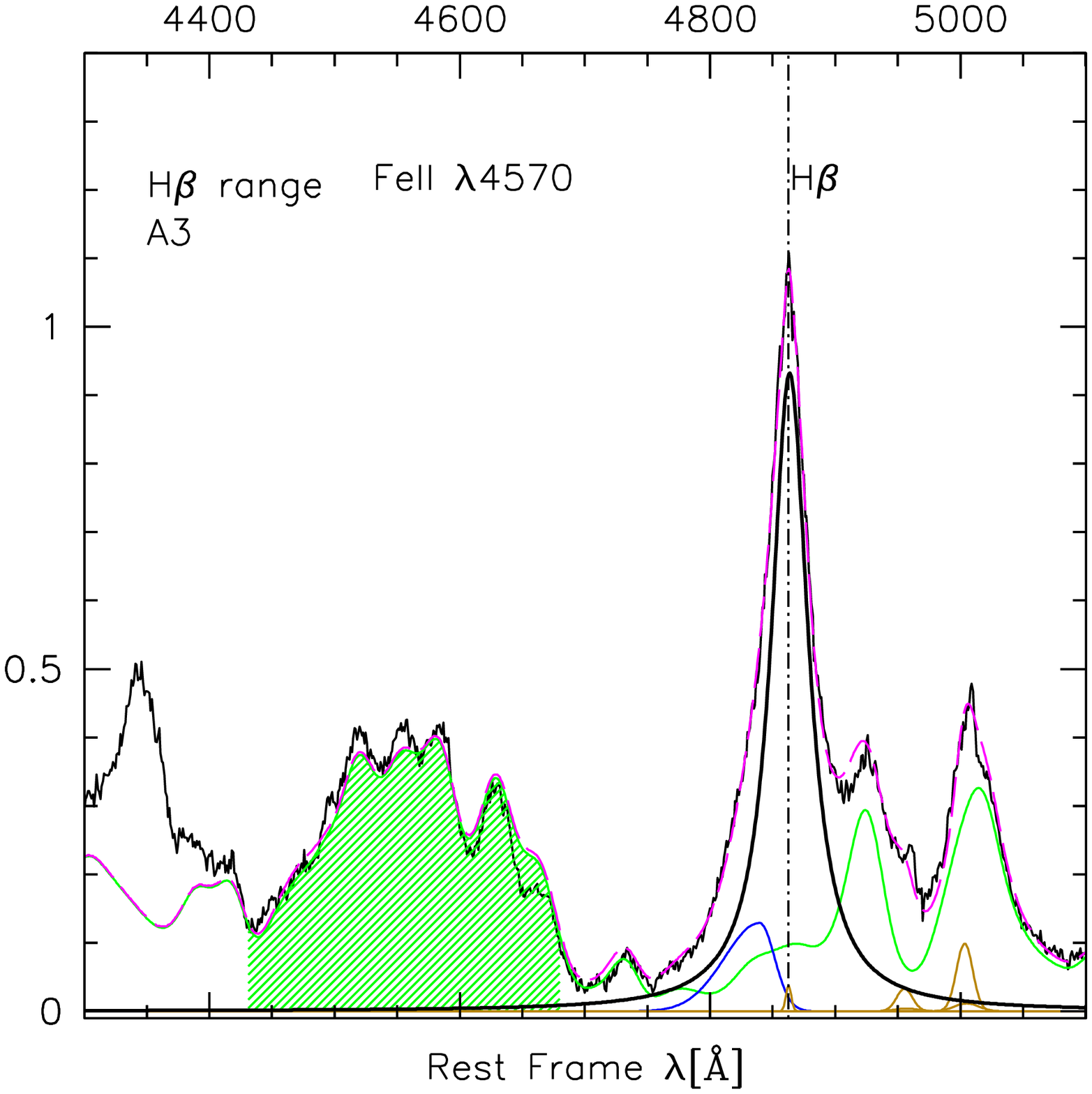}
\includegraphics[width=3.1in]{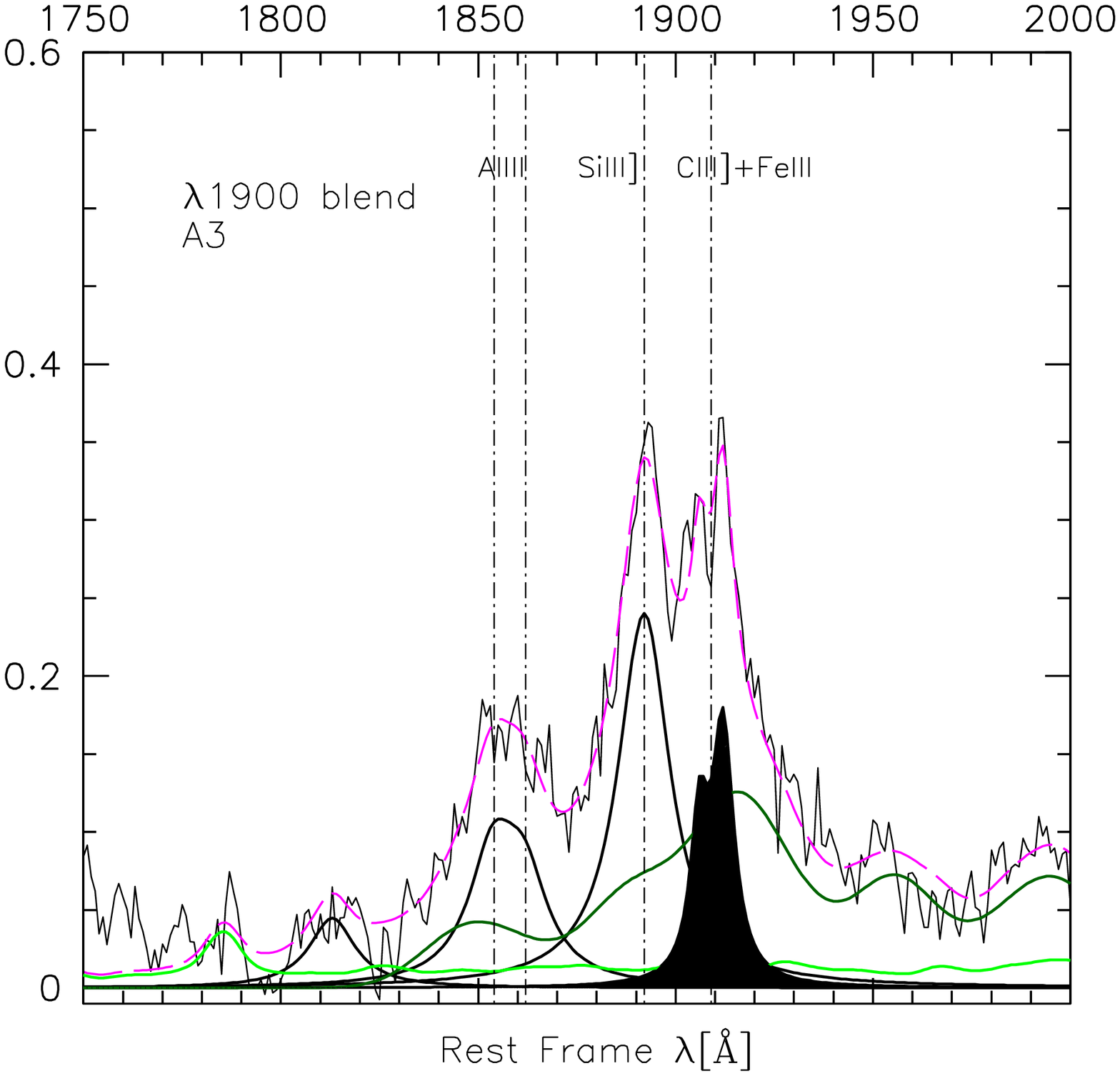}
\caption{Left panel: \hb\ spectral range for spectral type A3 i.e., for sources meeting the xA criterion \rfe$\ge1$. The spectrum is a median composite from the low-$z$ sample of \citep{marzianietal13a}. The \feiiq\ feature used to measure the \feiiopt\ prominence in the \rfe\ parameter is shown by  a lemon green shading. The thick Lorentzian line identifies the \hb\ broad component used as a virial broadening estimator, the lemon green line traces the adopted \feiiopt\ template, and the gold line the \oiiiopt\ emission. The   asymmetric residual  on the blue side of \hb\ represents an outflow contribution that is much more prominent in the \civ\ HIL line. Right panel: Spectral range of the 1900 blend for the same spectral type. The spectrum is, in this case, a median composite from the HST/FOS sample of \citep{bachevetal04}. The thick lines identify the \aliii\ and \siiii\ emission used to identify extreme sources. The shaded line is a blend of \ciii\ and \feiii\ $\lambda 1914$. The thick dark green line traces a template \feiii\  emission. \feii\ (lemon green line) is weak and practically negligible   in this range.  }
\label{fig:a3}
\end{figure}

\paragraph{At low $z$} Spectral types have been defined along the E1 sequence and extreme  quasars cluster in spectral types A3 and A4 (Fig. \ref{fig:e1f}).  The 4DE1 parameters for I Zw 1 (spectral type A3) are extreme, making it a candidate for the sources we are seeking.  \mbh\ and \lledd\ estimates for this source following the conventional methods described earlier  yield $\log$ \mbh\ $\approx 7.3 - 7.5$\ in solar units and $\log$ \lledd\  $\approx  -0.02 \pm0.11$ \citep{vestergaardpeterson06,negreteetal12,trakhtenbrotnetzer12}. The optical and UV emission line properties of I Zw 1 allow for the following empirical selection criteria: (a)  ratio of the intensity of \feii\ emission in the blend at 4570 and \hb, $R_\mathrm{FeII} > 0.5$ (at low $z$) and (b) \aliii/\siiii $\ge$ 0.5\ and \siiii/\ciii $\ge$ 1.0 (at high $z$). Fig. \ref{fig:a3} shows the \hb\ and 1900 spectral ranges for two composite A3 samples. The optical and UV selection criteria are equivalent \citep{marzianisulentic14}.

\begin{figure}[ht]
\begin{minipage}[t]{0.475\linewidth}
\centering
\includegraphics[width=3.25in]{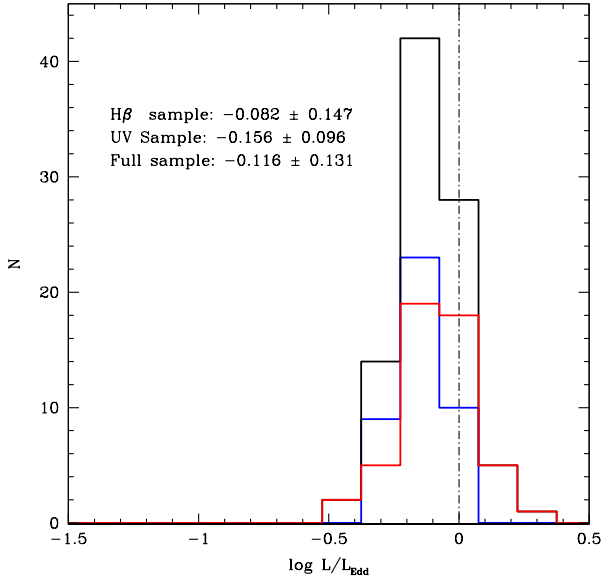}
\end{minipage}
\hspace{0.7cm}
\begin{minipage}[t]{0.475\linewidth}
\vspace{-7.75cm}
\centering
\caption{Distribution of Eddington ratios for the samples of highly accreting quasars identified in \cite{marzianisulentic14}. Thick line: full sample; blue line: UV sample with $2 \le z \le 2.6$; red-line: \hb\ sample. See text for more details. \label{fig:lledd}}
\end{minipage}
\end{figure}

These criteria were used to isolate a sample of extreme quasars from the SDSS DR4, mainly in the redshift ranges 0.4 -- 0.7 (applying criterion (a)), and 2.0 -- 2.7 (applying criterion (b)).  Fig. \ref{fig:lledd} shows the \lledd\ distribution that is tightly peaked around a limiting value with a relatively small scatter, $\approx$ 0.15 dex \cite{marzianisulentic14,marzianisulentic14a}.  The agreement between the samples selected using low-$z$ \hb\ and high-$z$ 1900 blend  is excellent: there is just a small systematic deviation   by about 20 \%. While this small deviation is not undermining the consistency of the optical and UV criteria, its amplitude is large enough to have a significant effect if xA sources are intended to be used as distance indicators for deriving cosmological parameters (\S \ref{cosmo}). 

Are   criteria (a) and (b) a sufficient condition for identifying xA sources? Apparently so, since we were unable to find sources satisfying these criteria for which there is a deviation from the condition \lledd\ = 1 that is significantly different from the one expected from the measurement errors convolved with the main known sources of uncertainty.  The criteria  are also expected to include the ``super-Eddington accreting massive black holes" (SEAMBHs; \cite{wangetal14a}) that show steep X ray spectra \cite{wangetal13}. The very same criteria may not however offer  a necessary condition: the emission-line ratios   employed as diagnostics are metallicity sensitive \citep{nagaoetal06}. Criteria (a) and (b) isolate sources that are heavily enriched (5 -- 10 $Z_\odot$), and it is at present unclear whether we can expect a tight relation between metallicity and \lledd\ \citep{shinetal13}. 

\paragraph{At high $z$ and $L$} Given a relation between emitting region radius and luminosity,  $r_{\rm BLR} = r_0 (L/L_0)^\alpha$,   the virial mass  can be rewritten for FWHM=$\Delta v$:  ${\rm FWHM} = c_1 \cdot (c_2)^{\frac{\alpha}{2}} L^{-\frac{\alpha}{2}} M_\mathrm{BH}^{\frac{1}{2}}$.  We can also write FWHM = FWHM$_0 L^{\frac{1-\alpha}{2}}  (L/L_\mathrm{Edd})^{-\frac{1}{2}}$, where the  $\alpha \approx 0.5$\  \citep{bentzetal09a}. The first equation  shows that the FWHM increases moderately with \mbh\ at a fixed luminosity. The effect is, albeit not first order,  noticeable on line widths since quasars span a broad \mbh\ range  ($ 10^6 -10^{9.5}$ \msol\ \citep{sulenticetal06,mcluredunlop04,woourry02a,sulenticetal12}). The second equation indicates that for a given \lledd,  the FWHM increases slowly with luminosity. In addition, there is a minimum FWHM at any $L$ if there is a defined maximum \lledd\ (Fig. \ref{fig:lum} adapted from \cite{marzianietal09}). In practice, for a limiting \lledd = 1\ the increase in FWHM as a function of luminosity is modest if $\log L \ltsim 46$ \  [\ergss], and significant above this limit. For $\alpha = 0.5$, there are no sub-Eddington sources with FWHM \hb\ lower than 3000 \kms\ at $\log L \ltsim 47$\ [\ergss]. Therefore we expect that sources radiating at the same Eddington ratio will show larger line width for the same \lledd. This simplified discussion emphasises the limitation introduced by the 2000 \kms\ criterion to identify NLSy1s. The NLSy1 prototype    I Zw 1 shows \lledd\ close to 1, as do {\em all} sources with \rfe$\gtsim$1. If we want to consider a {\em physical definition} of NLSy1s as sources radiating close to the Eddington limit, then it is necessary to generalize to broader line widths \cite{dultzinetal11}, employing a defining criterion that is independent of line width and instead traces  the source accretion state. 

\subsection{Spectral energy distribution}

Several analyses  consistently indicate that a slim accretion disk should  emit a steep soft and hard  X-ray spectrum \citep{szuszkiewiczetal96,wangnetzer03,cao09},  with hard X-ray photon index (computed between 2 and 20 KeV) converging toward $\Gamma_\mathrm{hard} \approx 2.5$ and bolometric  luminosity saturating to $L \approx \lambda_{\rm L} \left[ 1 + const.  \ln \left(\frac{\dot{m}}{50}\right)\right] M_\mathrm{BH}$,  \citep{mineshigeetal00}, and $\lambda_{\rm L}$\ is a constant related to the asymptotic $L/M_\mathrm{BH}$\ ratio for $\dot{m} \rightarrow \infty$.  

Observationally, the medium/high-energy  SED analysis is complicated by a soft X-ray excess and a hard X power-law extending to very high energy ($>$ 20 KeV).  The simplest interpretation of these SED components is optically-thick   (the soft X excess, \cite{bolleretal96,laoretal97b,grupe04}) and optically thin Comptonized radiation (the hard X ray power law), probably associated with a hot corona surrounding the black hole.  The soft-X excess is frequently detected in the case of highly accreting sources even if,  in very recent times, a few AGN have been discovered where the bare accretion disk may be the sole FUV soft-X emitter \cite[e.g.,][]{doneetal12}.  A wealth of physical information can be obtained from the soft-X and FUV domains.   However, the ionizing continuum SED and the associated  bolometric correction  of extreme quasars are still  poorly constrained  and a subject of current studies \cite[e.g., ][]{richardsetal06,shulletal12}. Even if there is a general consensus that xA quasars show a steep X-ray spectrum, it is not obvious how the far UV/soft X spectral region is affected by Comptonization of cold disk photons (a conventional explanation for the soft X ray excess),   by a disk temperature increase associated with a higher black hole spin (larger parameter $a$), and/or by the presence of a slim geometry at high accretion rate.

\begin{figure}[ht]
\begin{minipage}[t]{0.45\linewidth}
\centering
\includegraphics[width=2.1in]{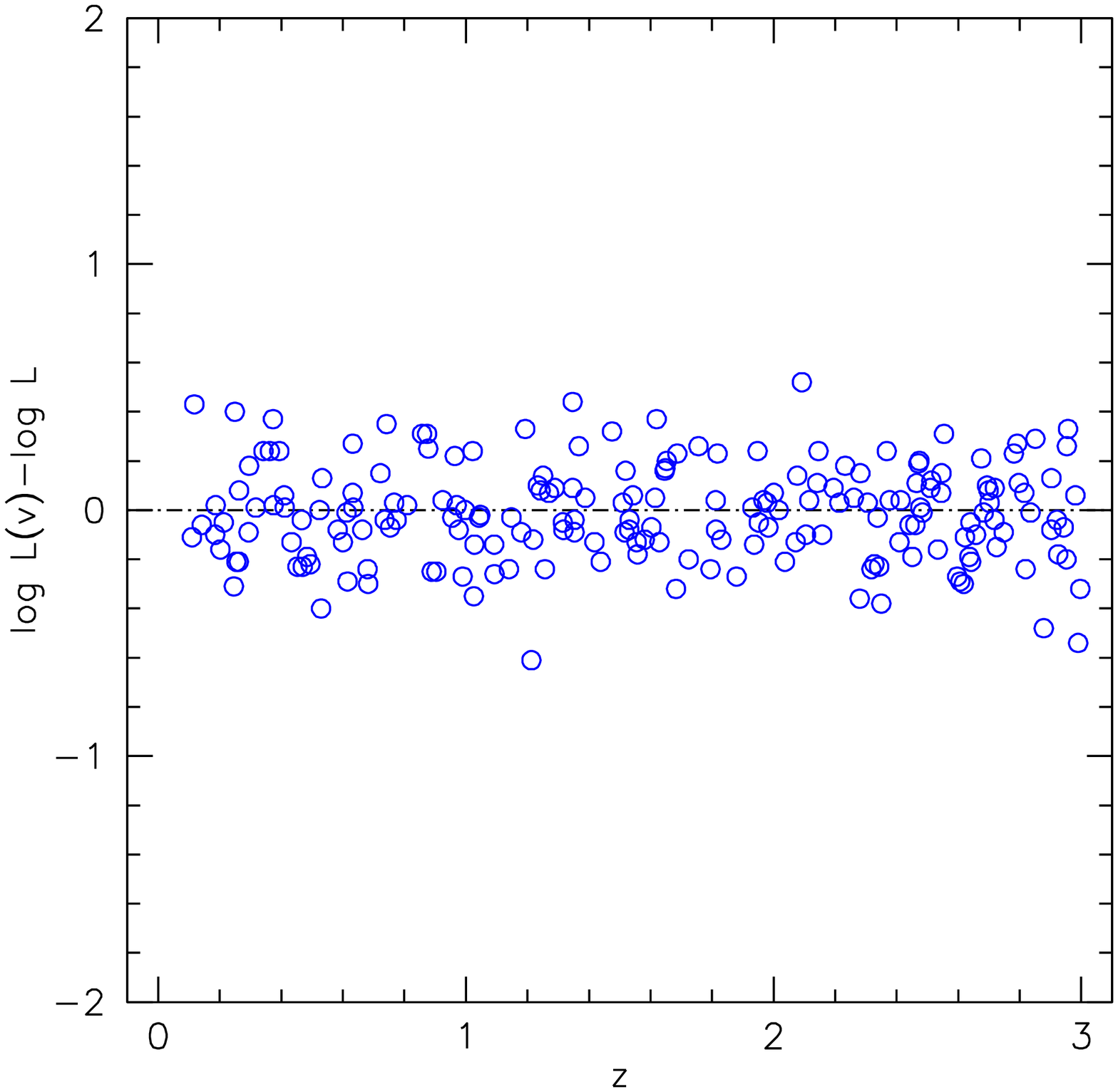}\\
\includegraphics[width=2.1in]{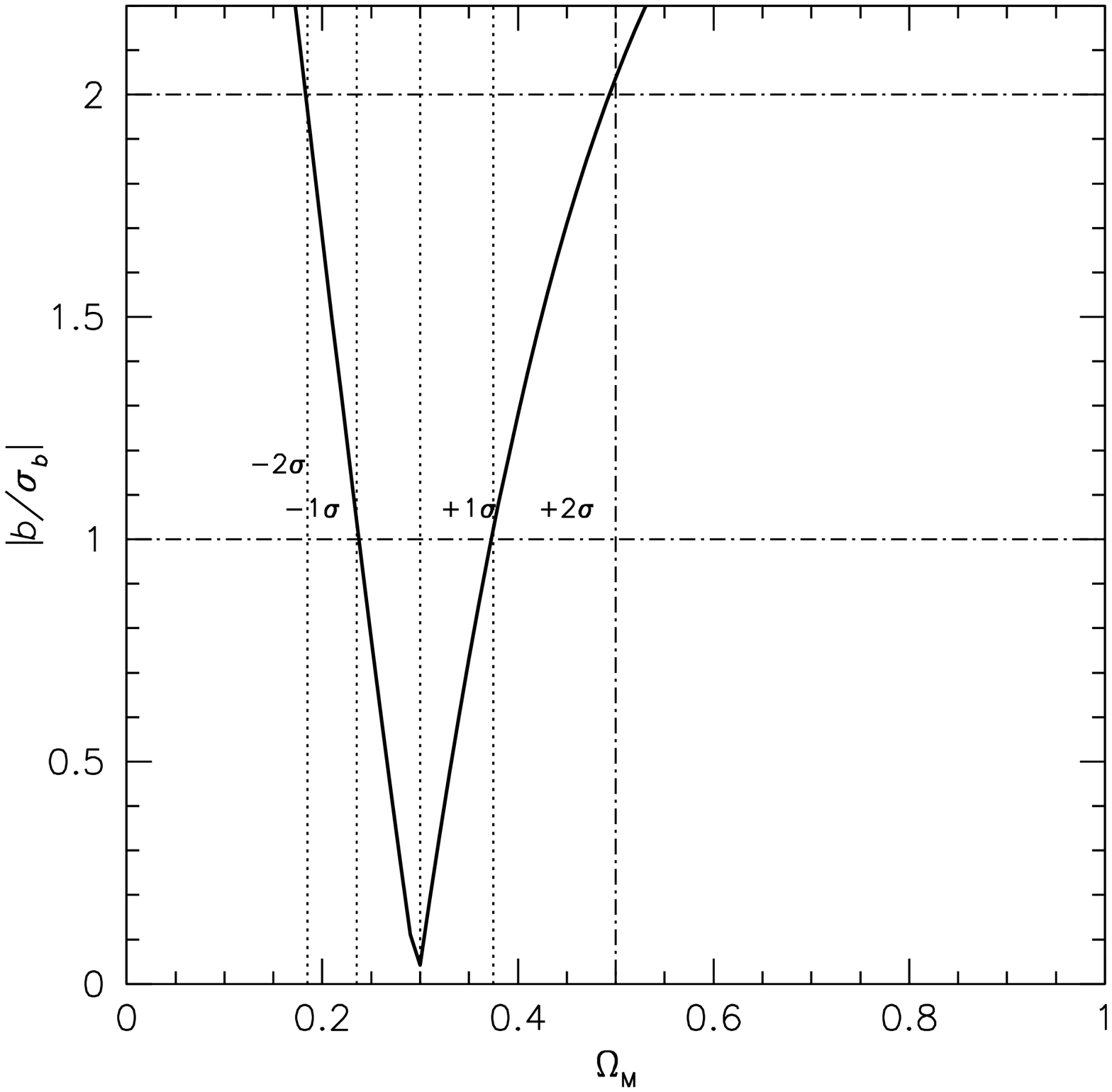}\\
\includegraphics[width=2.2in]{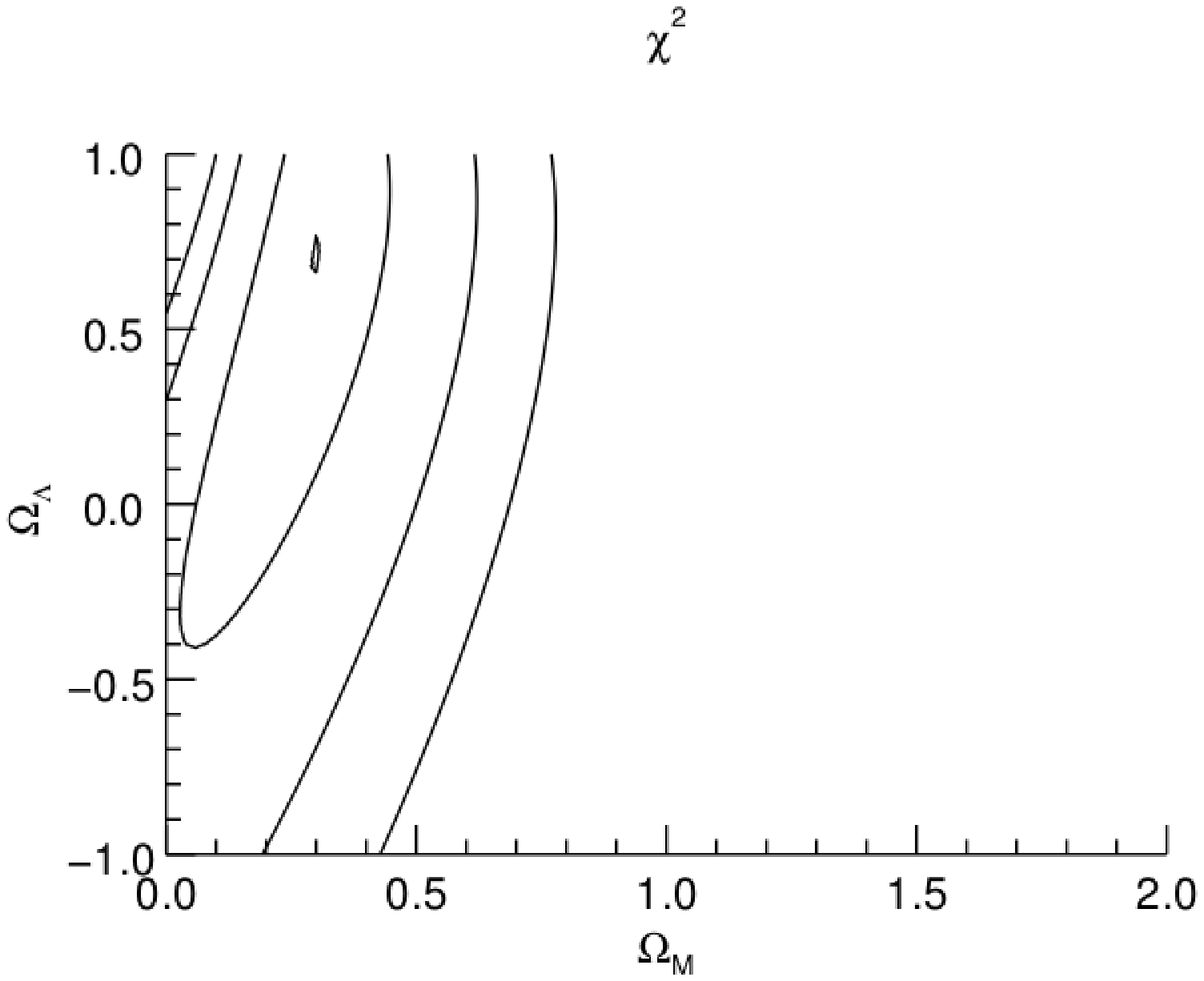}\\
\end{minipage}
\hspace{0.7cm}
\begin{minipage}[t]{0.5\linewidth}
\vspace{-5cm}
\centering
\caption{Results for a simulated sample of 200 highly accreting quasars with an rms scatter of 0.2 dex in luminosity around the concordance cosmology solution, and no systematic effects. Upper panel: behaviour of the luminosity residuals as a function of $z$. The luminosity residuals $\Delta \log L$\ are defined by the difference between the virial luminosity estimate (that is redshift independent) and the luminosity computed using the customary relation  linking it to the observed fluxes and to the luminosity distance $d_\mathrm{L} = d_\mathrm{L} (H_0,$\om, \ol, $z$); middle panel: behaviour of the ratio slope over slope standard deviation $b/\sigma_\mathrm{b}$ as a function of \om\ with the condition \om + \ol=1, for the same mock sample. The slope is computed fitting the residuals $\Delta \log L$($z$) with a linear function in the form $ a+ b \cdot z$. The ratio $b/\sigma_\mathrm{b}$ provides the best solution for $b \equiv 0$ and a confidence limit estimator.  The dotted lines identify the $\pm$ 1,2 $\sigma$\ confidence level. Lower panel: $\chi^2$\  in the plane \ol\ vs \om. Lines trace the 1,2,3 $\sigma$\ confidence levels. Note that a 3$\sigma$\ confidence limit upper limit to \om -- a remarkable feat in cosmology -- could be placed even with a relatively small sample. See \cite{marzianisulentic14,marzianisulentic14a} for a more detailed discussion.\label{fig:cosmo}  }
\end{minipage}
\end{figure}

\section{Highly accreting quasars: promising standard candles?}
\label{cosmo}

Quasars show properties that make them potential cosmological probes \citep[e.g,][]{marzianietal03d,bartelmannetal09}: they are plentiful, very luminous, and are detected at very early cosmic epochs (currently out to redshift 7).  However, they have never been successfully exploited as distance indicators in the past decades. Their  luminosity is spread over six dex, making them antithetical to conventional standard candles.  In addition, attempts at providing  one  or more parameters tightly correlated with   luminosity were largely unsuccessful: the ``Baldwin effect" did not live up to its cosmological expectations, \cite{marzianietal08,bianetal12}. 

Several research groups \cite{wangetal13,wangetal14a,watsonetal11,lafrancaetal14,czernyetal13,kingetal14,marzianisulentic14} are now focusing on the use of quasars to constrain two main cosmological parameters: the energy density of the Universe associated with matter and the cosmological constant. Two main different approaches  are proposed, as summarily reviewed by \cite{marzianisulentic14a}: the use of ``standard rulers'' (basically, the linear size of the broad line emitting region, \cite{elviskarovska02,watsonetal11}) or luminosity estimates independent of redshift.  

Realistic expectations are now kindled by  isolating a class of quasars with some constant property from which the quasar luminosity can be estimated independently of its redshift.  As mentioned earlier, in super Eddington accretion regime, a geometrically and optically thick structure known as a ``slim disk" is expected to develop \cite{abramowiczetal88}.  Quasars hosting slim disks should radiate at a well defined limit because their luminosity is expected to saturate close to the Eddington luminosity even if the accretion rate becomes highly super-Eddington. 


If \lledd\  can be derived from some distance-independent measure, it becomes possible to derive distance-independent quasar luminosities.  The  Eddington ratio is proportional to the ratio of source luminosity to black hole mass.   A virial broadening estimator can be extracted from the FWHM or velocity dispersion of  suitable optical and UV lines \citep{marzianisulentic12,shen13,negreteetal13}.  \citet{negreteetal13} derived a relation between $ r_\mathrm{em} $ and  the ionizing photon flux $\Phi$\ illuminating the line emitting region. This relation is based on diagnostic ratios and is fully independent of  $z$-derived parameters.  The basic equation relating line width to luminosity (the ``virial luminosity equation'') can then be written as: $L \propto$ (\lledd)$^2 \kappa \Phi^{-1} (\delta v)^4$, where $\kappa$\ is a function of quasar intrinsic properties  (see \cite{marzianisulentic14} Eq. 6 for the derivation of a fully developed expression; also cf. Eqs. 6 and 7  of \citep{lafrancaetal14}).  

The advantage of  extreme radiators is that (1) a large sample of highly accreting quasars can be selected with relative ease from recent major surveys (i.e. SDSS IV -- BOSS); (2) the sample will cover a broad redshift range, up to $z \approx 3 - 4$; (3) the redshift coverage will be uniform, save for a small gap at $z \approx 0.8 - 1.1$\ due to atmospheric absorption; (4) selection criteria are empirically defined and do not require model-dependent assumptions.  Quasars can  provide a fully independent measure from supernovae, baryonic acoustic oscillations, and clusters of galaxies that suffer from a rather low redshift detection limit ($z \approx 1$). Quasars can be easily detected up to $z \approx 4 $ but those in the range $1 < z < 3-4 $\ are of greatest interest because the effect of the cosmic matter density is believed to dominate over the repulsive effect of the cosmological constant. Extremely accreting quasars can yield an independent measure of $\Omega_\mathrm{M}$\ with tight limits. 

xA quasars luminosity estimates  from the virial equation are still plagued by a relatively large scatter. Several observational and physical aspects of xA sources should be clarified before taking the leap of a cosmological application.   The analysis of \citet{marzianisulentic14} indicates that a major source of statistical error  is the scatter due to  orientation effects. Line widths  depend on  viewing angle, with a maximum  effect estimated to be about a factor 2 (e. g., \cite{collinetal06}). An estimate of the viewing angle can be obtained through disk wind models \citep{dotanshaviv11,proga03} of the \civ\ line whose profile is affected by a strong radial component associated with outflowing gas  \cite{murraychiang97,flohicetal12}. 

In addition to a reduction of the rms associated with an increase in sample size,  it is necessary also to ensure a minimum S/N: measurements on  mock emission line blends indicate that a S/N $\approx$ 20 could yield virial broadening estimators with a statistical accuracy of $\approx$ 10\%\ for W$\approx$5 \AA.  Mock samples   indicate that a large number ($\sim$ 1000) and   wide redshift coverage should allow us to measure $\Omega_\mathrm{M}$ with a precision $\delta \Omega_\mathrm{M}  \approx \pm$ 0.1 at a 2 $\sigma$\ confidence level \cite{marzianisulentic14}, and to meaningfully constrain $\Omega_\Lambda$. 

A large sample, uniformly distributed in redshift would make it possible  to consider the evolution of the equation of state of dark energy as a function of redshift.  Current issues go beyond the existence of the dark energy and focus more on its properties. The simplest model for dark energy is a cosmological constant with a fixed equation of state ($p = w \rho$, with fixed $w=-1$). However, the dark energy density may depend weakly upon time, according to many proposed models of its nature \cite{carroll05}: a general scalar field predicts $w$\ to be negative and evolving with redshift. The strength of an xA sample is the ability to build the Hubble diagram covering uniformly a broad range of redshifts, with the expected precision scaling with the inverse of the square root of the number of sources. xA quasars may turn out to be ideal distance indicators to test whether the dark energy equation of state is constant or is evolving as function of redshift (following selected parametric forms for $w(z)$, \cite{capozzielloetal14a,capozzielloetal14b}).





\section{Conclusion}

We have here reviewed recent results obtained especially in \citep{marzianisulentic14,marzianisulentic14a,sulenticetal14b,sulenticetal14,sulenticetal14a}. Highly accreting quasars can be identified  and studied most easily in the 4DE1 context originally proposed by \citet{sulenticetal00a}. Since the ``main sequence'' in the parameter space is primarily governed by \lledd, highly  accreting sources cluster at one end of the quasar distribution. It is possible to define  a sufficient condition for  identifying highly accreting quasars that can be applied to large samples of spectra. The same criteria may also work for RL sources. There is   a straightforward explanation of the Population A/B dichotomy in terms of structure of the accretion flow, and highly accreting quasars are most likely the ones radiating at the highest accretion rate, with luminosity saturating toward an asymptotic $L/M$\ ratio. If so,   xA quasars offer unprecedented possibilities as distance indicators. 

Many open issues concern xA sources. We still now relatively little about their SEDs, their host properties, their emission line properties in the \hb\ range at high $z$.  Physically, there is no established connection between the broad line emitting region structure and the presence of a thick structure. The use of xA sources as ``Eddington standard candles" is really a fascinating opportunity whose practical fullfilment  rests however on  obtaining a clear view of the broad line emitting region structure.

\begin{multicols}{2}\small
\bibliographystyle{apj} 

\expandafter\ifx\csname natexlab\endcsname\relax\def\natexlab#1{#1}\fi
\end{multicols}

\end{document}